**Title:** Gray Matter Volume Correlates of Co-occurring Depression in Autism Spectrum Disorder

Running Title: rGMV Correlates of co-morbid depression in ASD

Authors: Dolcy Dhar[1#], Manasi Chaturvedi[1,2#], Saanvi Sehwag[1], Chehak Malhotra[3], Udit[4], Chetan Saraf[5], Mrinmoy Chakrabarty[1,2#*]

[1]Dept. of Social Sciences and Humanities, Indraprastha Institute of Information Technology Delhi, New Delhi 110020, INDIA.

[2]Centre for Design and New Media, Indraprastha Institute of Information Technology Delhi, New Delhi 110020, INDIA.

[3]Dept. of Mathematics, Indraprastha Institute of Information Technology Delhi, New Delhi 110020, INDIA.

[4]Dept. of Computational Biology, Indraprastha Institute of Information Technology Delhi, New Delhi 110020, INDIA.

[5]Dept. of Computer Science and Engineering, Indraprastha Institute of Information Technology Delhi, New Delhi 110020, INDIA.

**# equal contribution (Co-first author)**

Correspondence (*):

Mrinmoy Chakrabarty, Ph.D.

Assistant Professor (Cognitive Science), Dept. of Social Sciences and Humanities, Indraprastha Institute of Information Technology Delhi (IIITD), New Delhi 110020, INDIA. TEL: +91-011-26907-363; E-mail: mrinmoy@iiitd.ac.in

# Abstract

**Background**

Autism Spectrum Disorder (ASD) involves neurodevelopmental syndromes with significant deficits in communication, motor behaviors, and emotional and social comprehension. Often, individuals with ASD exhibit co-occurring depression characterized by a change in mood and diminished interest in previously enjoyable activities. Due to communicative challenges and a lack of appropriate assessments in this cohort, co-occurring depression can often go undiagnosed during routine clinical examinations and, thus, its management. The literature on co-occurring depression in adults with ASD is limited. Therefore, understanding the neural basis of the co-occurring psychopathology of depression in ASD is crucial for identifying brain-based markers for its timely and effective management.

**Methods**

Using structural MRI and phenotypic data from the Autism Brain Imaging Data Exchange (ABIDE II) repository, we examined the pattern of relationship regional grey matter volume (rGMV) has with co-occurring depression and autism severity within regions of a priori interest in adults with ASD (n = 44). Further, we performed a whole-brain exploratory analysis of the rGMV differences between ASD and matched typically developed (TD, n = 39) samples.

**Results**

The severity of co-occurring depression correlated negatively with the rGMV of the right thalamus. Additionally, a significant interaction was evident between the severity of co-occurring depression × core ASD symptoms towards explaining the rGMV in the left cerebellum crus II. The whole-brain exploratory analysis remained inconclusive.

**Conclusion**

The results further the understanding of the neurobiological underpinnings of co-occurring depression in adults with ASD and are relevant in exploring structural neuroimaging-based biomarkers in the same cohort.

**Introduction**

Autism spectrum disorder (ASD) is a set of neurodevelopmental syndromes that exhibits a diverse range of symptoms marked by profound social communication and interaction deficits, often leading to an inability to comprehend others' emotions and reciprocate social cues, often accompanied by a distinctive pattern of repetitive and restrictive behavior/interests(1).

Apart from the core deficits, individuals with ASD often exhibit co-occurrence or the co-occurrence of two or more disorders in the same individual, e.g., epilepsy, gastrointestinal symptoms, metabolic and immune disorders, sleep and feeding problems, and psychiatric disorders (2–4). The psychiatric comorbidities in individuals with ASD spanning childhood through adulthood tend to be high (5–10). Amongst the psychiatric co-occurring, attention deficit hyperactivity disorder, obsessive-compulsive disorder, anxiety, and depression are reported frequently (10–15). Co-occurring, including depression, leads to further aggravation of daily functioning in individuals with ASD, aside from increasing the treatment costs. Besides, it is challenging to diagnose co-occurring psychopathologies due to communication deficits, atypical manifestations, and paucity of diagnostic screens for these disorders in individuals with ASD (2,16). Hence, co-occurring psychopathologies frequently remain undiagnosed or under-recognized until the adult stage and are also undertreated in ASD, one of the prominent ones being depression (10,12) While an earlier meta-analysis places the rate of lifetime depression at 37% in adults with ASD (17,18) Given the considerable prevalence of co-occurring depression in ASD, a better understanding of the neurobiological bases of the same is of merit as objective, brain-based markers have the potential to aid in better identification and characterization of co-occurring depression. This is especially relevant for adults with ASD, in whom the diagnosis is even more challenging owing to the individual's compromised abilities to recognize emotions/thoughts and interact emotionally with others. The picture is further clouded by the atypical presentations of depression in ASD and the overshadowing of salient depressive symptoms by the core deficits of ASD(16,19). Thus, brain-based insights here may have important implications for identification, time course of illness, and the overall management of the various issues presenting in adults with ASD and co-occurring depression, as there could be shared biological mechanisms between the core manifestations of ASD and its co-occurring depressive symptoms.

Depression is an etiologically heterogeneous condition and primarily a disorder of the representation and regulation of mood and emotion. It is characterized by depressed mood, diminished interests, disturbed sleep and appetite, fatigue or loss of energy, as well as impaired cognitive functions (20–22). Studies reporting structural brain differences in depression implicate a few brain regions - caudate, putamen, globus pallidus, thalamus, hippocampus, anterior cingulate cortex, and orbitofrontal cortex (21,23). The brain regions of the striatum (caudate, putamen, and globus pallidus) are part of the mesolimbic dopaminergic system and have been associated with decreased motivation, diminished interests plus energy levels of individuals with depression, and reported to contribute significantly to the expression of depressive symptoms(24,25) Similarly, orbitofrontal cortex in the frontal region of the brain is a crucial structure for valuation of rewards, mood, and emotion which are affected in depression (26) In the subcortical limbic region of the brain, thalamus is known to control mood in depression owing to its connections with dorsal/ventral prefrontal cortex and amygdala (27). Also, the hippocampus in the subcortical region of the brain is involved with learning and memory (episodic, declarative, spatial, and contextual), which are often compromised in depression (21,28). Additionally, an increasing body of evidence now suggests that the cerebellum, in addition to its role in motor function, also has a role in emotion regulation and its disorders, e.g., depression (29)

The literature is replete with information on structural anomalies of different regions of the brain in ASD. However, the focus predominantly has been on the structural differences between ASD vs. TD and the relationship between these structural anomalies and the severity of the core symptoms of ASD in children and adolescents (30–39). Conspicuously, co-occurring depression in adults with ASD and its relation to the structural brain changes towards explaining the severity of core symptoms of ASD is relatively less attended in the extant literature, to our knowledge. Additionally, in neuroimaging studies of co-occurring depression and ASD, previous research has utilized alternative imaging modalities, such as resting-state fMRI, predominantly focusing on children and adolescents(40–42). Thus, despite the literature and evidence of the presence of co-occurring depression in individuals with ASD, the investigation of structural biomarkers for the same remains understudied. Developmental trajectories and manifestations of autism are more varied in adulthood and further clouded by symptoms of co-occurring depression, making clinical assessment challenging. Given the above, we explored the link across co-occurring depression – brain structural differences – the severity of core ASD symptoms in this study.

Specifically, we employed region-based morphometry of regional grey matter volume (GMV) on structural T1 magnetic resonance imaging (MRI) data of adults (18 – 35 years) from the Autism Brain Imaging Data Exchange II repository(43) to investigate the following in a priori regions of interest in the brain - ROIs - a) association between the severity of co-occurring depression in ASD and their respective rGMV); b) association between the interaction of the severity of co-occurring depression × core ASD symptoms and rGMV in ASD. We chose the ROIs - anterior cingulate cortex, thalamus, and cerebellum as they pertained to some salient issues of social cognition, sensory-motor function, motor coordination, theory of mind, and emotion that were relevant in both individuals with ASD and depression. The rationale behind choosing and the steps towards defining the ROIs are detailed in the Supplement. Aside from the above two, we investigated the whole brain differences of rGMV between individuals with ASD and their matched TD peers in an exploratory manner with and without the severity of co-occurring depression to look into effects outside the domain of our a priori interest.

## Methods

*Participants*

We sourced all structural T1-weighted MRI data of individuals aged 18 – 35 years with ASD and matched TD from the Autism Imaging Data Exchange (ABIDE II) database used for the analyses (43). ABIDE-II was selected as ABIDE-II's dataset, which placed a stronger emphasis on core ASD symptoms and closely adhered to the DSM-5 diagnostic criteria.

Furthermore, the sites included in ABIDE-II closely matched our inclusion criteria in terms of age and diagnostic assessment, enhancing the relevance of our research to the target population. Conversely, ABIDE-I posed challenges due to missing data in the phenotypic files essential to our study. Anonymized data in the adult age bracket that contained scores of Autism Diagnostic Observation Schedule (ADOS) in ASD and Beck Depression Inventory (BDI) for both ASD and TD were only included in the study. We used the research–reliable ADOS-2 score, which is a well-established metric for assessing the clinical severity of autism(44), and the BDI score as a metric for the severity of depressive symptoms (45). Initially, we shortlisted 57 ASD and 50 TD individuals' data. A total of 18 (ASD = 10; TD = 8) scans were excluded on visual inspection for quality distortions and motion artifacts, as explained elsewhere(46). Five scans were excluded after the pre-processing step of the VBM analysis pipeline (described below) as their Image Quality Rating (IQR) was below 2.5 standard deviations from the mean in the sample. Consequently, we reached a final sample size of 44 ASD (mean ± SD: age = 21.23 ± 3.17 years, ADOS = 9.77 ± 2.83, BDI = 9.51 ± 9.55; 4 females) and 39 TD scans (mean ± SD: age = 22.36 ± 3.26 years; BDI = 5.89 ± 4.23; 9 females). Image acquisition parameters and the number of ASD and TD individuals for each scanning site are in Table 1. The steps of data pre-processing and defining the ROIs (Fig.1) a priori is in the Supplement. The study and its procedures were approved by the Institutional Ethics/Review Board of the Indraprastha Institute of Information Technology Delhi, INDIA.

**Table 1**: Summary of MRI scanning sites and parameters

| Scanning sites | Scanning parameters | Number of Participants | | Age of Participants (Avg. in years) | |
|---|---|---|---|---|---|
| | | ASD | TD | ASD | TD |
| Barrow's Neurological Institute | 3 T Ingenia's 15 channel Scanner; Sagittal MP-RAGE (TE = 'shortest'; TI = 900; FA = 9°; 170 slices) | 13 | 10 | 20.7 | 20.8 |
| Indiana University | 3 T Siemens Magnetom; Sagittal MP-RAGE (TE = 2.3; TR = 2400; TI = 1000ms; FA = 8°; 256 slices) | 15 | 15 | 21.6 | 22.3 |
| Olin Neuropsychiatry Research Centre | 3 T Siemens Magnetom; Transversal MP-RAGE (TE = 2.88; TR = 2200; TI = 794; FA = 13°; 208 slices) | 16 | 14 | 21.25 | 23.5 |

**Figure 1**

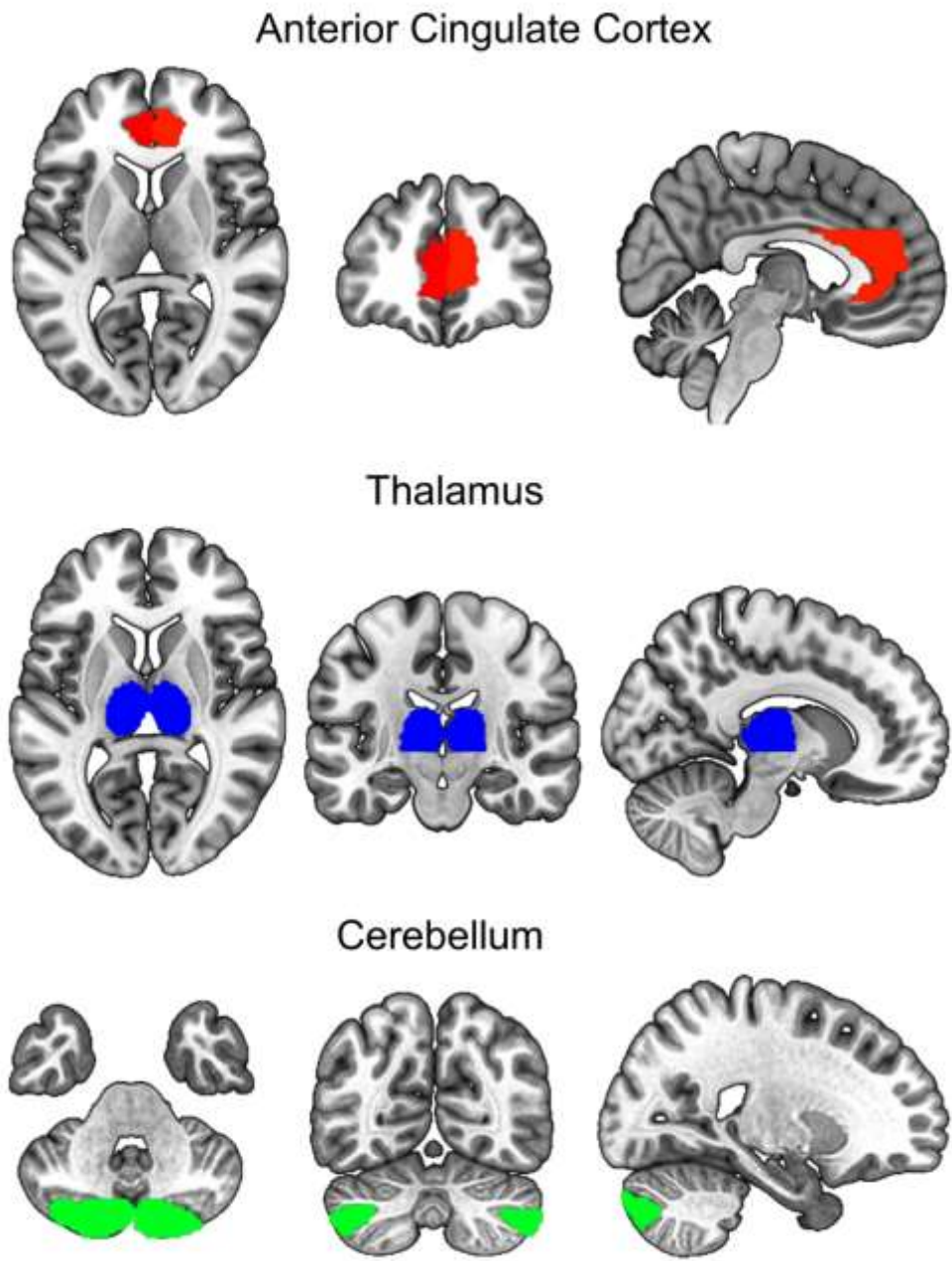

Fig. 1 The regions of a priori interest (ROIs) overlaid on a high-resolution T1 weighted template image from 152 normal subjects (MRIcron: MNI152.nii) for visualization. Axial, Coronal and Sagittal orientations (from left to right) of the same image and respective ROIs marked in different colors.

*Statistical analyses*

Within the ASD group, analyses of the a priori ROIs were done using the general linear model with rGMV of the brain as the dependent variable and total intracranial volume (TIV), age, scanner site, and BDI × ADOS (interaction term) as the independent variables. Thus, for each ROI, we used a multiple linear regression model of the form…

$$y = \beta_0 + \beta_1 X_1 + \beta_2 X_2 + \beta_3 (X_1 * X_2) + \beta_4 X_3 + \beta_5 X_4 + \beta_6 X_5 + \epsilon \quad (1)$$

where $y$ = rGMV of the ROI; $\beta_0$ = intercept; $X_1$ = BDI; $X_2$ = ADOS; $X_1 * X_2$ = interaction term between BDI × ADOS; $X_3$ = age; $X_4$ = scanner site; $X_5$ = TIV; $\beta_{1-6}$ = parameter estimates; $\epsilon$ = residual. We used age, scanner site, and TIV as covariates of no interest. The independent variables in the model were orthogonal. For the ROI analyses, we additionally set the significance level in each hemisphere using the Bonferroni correction described earlier (47–49). The total number of ROIs was three; thus, the significance level was set at $p \leq 0.05/3 \approx 0.02$ in each hemisphere.

On the other hand, for the between-groups (ASD vs. TD) exploratory voxel-wise whole-brain analysis of the rGMV, the independent variables were scanner site, TIV, and diagnosis. While the scanner site was included as covariates of no interest, age and gender were not as they were matched between the two groups (see Table 1). The TIV variable was not orthogonal to the dependent variable and, hence, excluded from the covariates. Then, the entire model was globally scaled with respect to their total intracranial volumes. The above analysis was done using the Computational Anatomy Toolbox v12.8.1 (r2043) (50) hosted on MATLAB 2022a. All thresholded statistical maps were visualized using MRICroGL (www.nitrc.org/projects/mricron)(51), and the brain regions were labeled using CAT12_AAL3 atlas(52).

**Table 2:** Demographic information of the subjects in two groups.

| Particulars | ASD ratio/mean (range) | TD ratio/mean (range) | Test statistic (pdf) | p-value | *Effect size* |
|---|---|---|---|---|---|
| Gender ratio (male : female) | 10:1 (ratio) | 10:3 (ratio) | $X^2_{(1)} = 0.84$ | 0.36 | Phi = 0.19 |
| Age in years | 21.23 (17-28) | 22.36 (18-31) | $t_{(39)} = -1.79$ | 0.081 | Cohen's *d* = 0.28 |
| BDI | 9.51 (0-33) | 5.89 (0-19) | $t_{(37)} = -2.36$ | 0.023 | Cohen's *d* = 0.38 |
| ADOS | 9.77 (4-18) | - | | | |

BDI = Beck's Depression Inventory; ADOS = Autism Diagnostic Observation Scale

## Results

*Relationship of rGMV with co-occurring depression and symptomatic severity within ROIs in ASD*

The set of ROIs of a priori interest (Fig.1) that were defined based on earlier reports (see Supplement for details) were then employed in multiple regression analyses to ascertain the association between the rGMV of each ROI with the severity of co-occurring depression (BDI) and the interaction between co-occurring depression × core ASD deficit (ADOS) severities respectively, after controlling for covariates - age, scanner site, TIV in the ASD sample (see details in *Statistical analyses*). Please note that while both the clusters reported below survive family-wise error rate (FWE) correction within their respective extents, they are at the edge of the statistical significance threshold after imposing an additional stringent correction for the three ROIs tested, as described in the section – *Statistical analyses*.

First, it revealed a significant negative correlation ($r = -0.54$, 95% confidence interval = [- 0.72, - 0.29]) between the rGMV and BDI in the right thalamus after the adjustment of $p$ value within the search volume ($t_{(39)} = 3.40$, $pFWE = 0.024$, cluster size $kE = 49$; ROI volume = 2,309 voxels of size 1.5 × 1.5 × 1.5 mm. Table 3-A, Figure 2), showing a systematic decrease in rGMV with increments of co-occurring depression severity in individuals with ASD.

Second, to clarify the relationship between rGMV and the interaction term between BDI × ADOS, the data from the ASD sample was initially median split into two levels based on the severity of core ASD deficits (ADOS score). Thus, two subsets were created - individuals with high ADOS (> 10 (median); mean ± s.e.m, 12.75 ± 1.94) and those with low ADOS scores (≤ 10; 8.07 ± 1.53). The goal was to tease out the existing trend in the relationship between BDI and rGMV at two different levels of ASD clinical severity indexed by ADOS. The interaction term was derived using the 'Interaction (Linear)' feature of the Computational Anatomy Toolbox v12.8.1 (r2043). The analysis yielded a significant interaction in the left cerebellum crus II after the adjustment of *p*-value within the search volume ($F_{(1,37)}$ = 15.37, *pFWE* = 0.019, *kE* = 66; ROI volume = 4,359 voxels of size 1.5 × 1.5 × 1.5 mm. Table 3-B, Figure 3). Here, a mutually opposite trend in the relationship was evident such that the rGMV and co-occurring depression severity varied by the symptomatic severity of core deficits (low and high ADOS) in individuals with ASD. While the rGMV showed a trend of decrement with increasing co-occurring depression in the individuals who are lower on the autism spectrum (*r* = -0.09, 95% confidence interval = [-0.49, 0.33]), it proportionately increased with increasing co-occuring depression in those who are relatively higher on the spectrum (*r* = 0.29, 95% confidence interval = [- 0.16, 0.64]).

*Exploratory analysis of the whole brain voxel-wise differences of rGMV between ASD and TD*

An independent sample *t*-test was done between ASD and TD, controlling for scanner site and BDI. The entire model was globally scaled with respect to their total TIV to account for variations in brain sizes, as the TIV variable was not orthogonal to the dependent variable. This revealed no significant differences between the two groups after correcting for family-wise error rate (FWER). Thus, we determined the rGMV differences using a more liberal cluster-forming threshold of $p < 0.001$ (uncorrected) and extent thresholds set at the expected number of voxels as suggested (50). Here, the left medial superior frontal gyrus (MSFG) showed a lesser rGMV in ASD than TD (Table 3-C, Figure 4A).

Next, an interaction analysis was done between the variables of diagnosis (ASD vs.TD) × depression (BDI) after controlling for the scanner site, which showed the rGMV in the left superior parietal gyrus (SPG) varying between the two groups (ASD vs.TD) at a threshold of $p < 0.001$ (uncorrected). While there was a trend of progressive decrease in the rGMV of SPG in TD individuals, there was a proportional increase in the rGMV of the same region with increasing severities of depression (BDI) in individuals with ASD. Details are in Table 3-D and illustration in Figure 4B.

**Table 3:** Results of statistical analyses

| Analysis | Brain region | Peak MNI coordinates | | | Cluster size (# voxels) | Cluster-level $p$-value |
|---|---|---|---|---|---|---|
| | | X | Y | Z | | |
| A) rGMV vs. BDI regression in ASD | Right thalamus | 18 | -22 | 14 | 49 | $p_{FWE}$ = 0.024 |
| B) rGMV vs. (ADOS x BDI interaction) in ASD | Left Crus II Cerebellum | -15 | -84 | -39 | 66 | $p_{FWE}$ = 0.019 |
| C) ASD<TD controlling for BDI | Left medial superior frontal gyrus | -10 | 30 | 42 | 116 | uncorrected $p < 0.001$ |
| D) Group (ASD, TD) x BDI interaction | Left superior parietal gyrus | -20 | -57 | -60 | 95 | uncorrected $p < 0.001$ |

Brain regions are labelled as per the AAL3 atlas in CAT12; $p_{FWE}$ Family-wise error rate corrected $p$-value

**Figure 2**

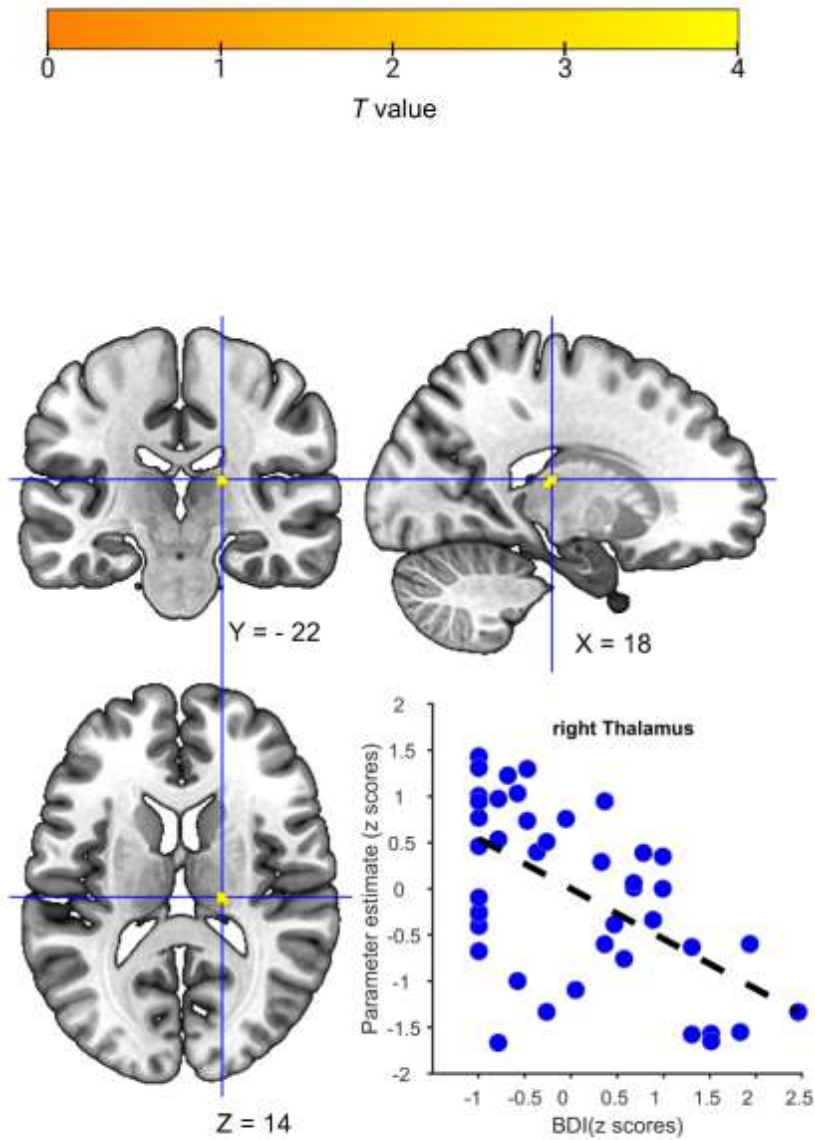

Fig. 2 Negative correlation between co-occurring depression severity (BDI) and rGMV (parameter estimate) in adults with ASD in the right thalamus. Thresholded statistical map ($p_{FWE} < 0.05$) overlaid on a high-resolution 1.5 T template image from a single normal subject (MRIcron: ch2.nii.gz) for visualization. Sagittal (X), Coronal (Y), Axial (Z) orientations of the same image and respective peak coordinates.

**Figure 3**

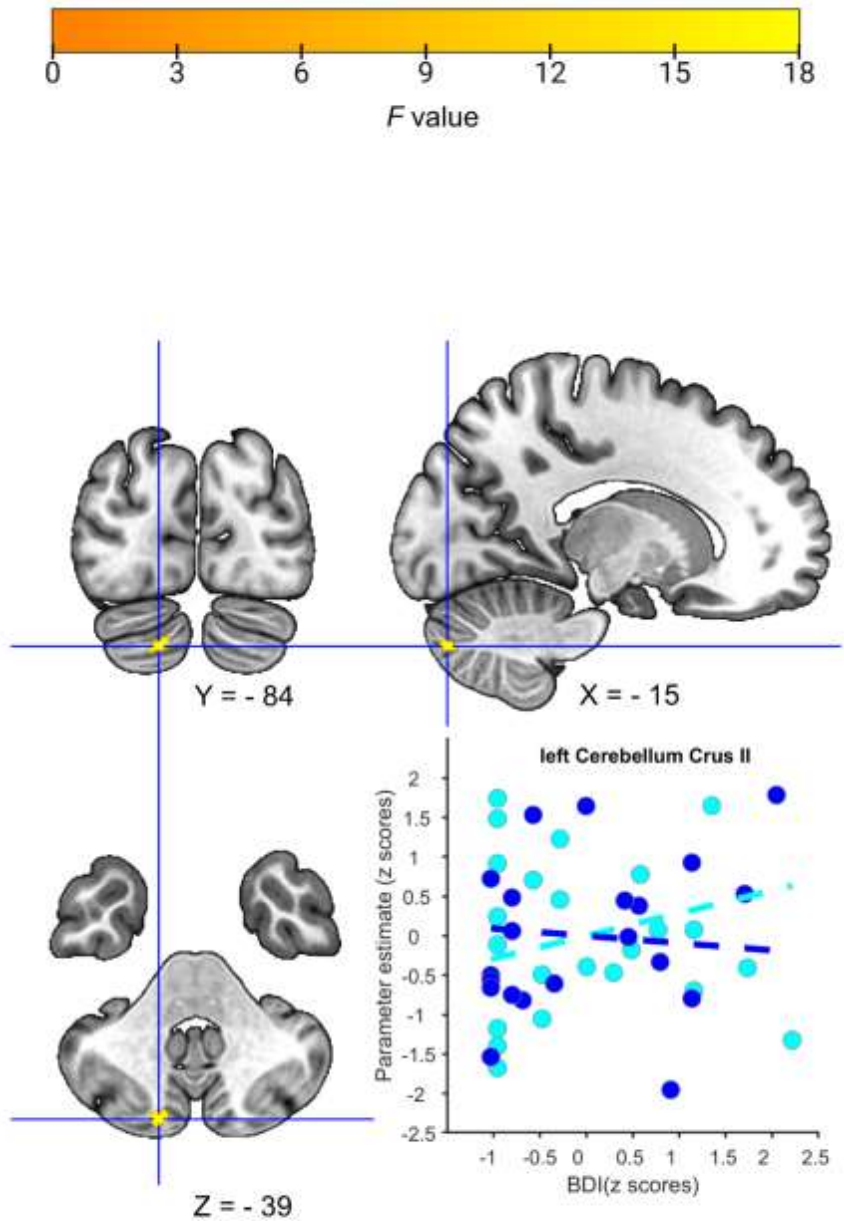

Fig. 3 Association between BDI × ADOS interaction term and rGMV (parameter estimate) in adults with ASD in the left cerebellum crus II. Thresholded statistical map ($p_{FWE} < 0.05$) overlaid on a high-resolution 1.5 T template image from a single normal subject (MRIcron: ch2.nii.gz) for visualization. Sagittal (X), Coronal (Y), Axial (Z) orientations of the same image and respective peak coordinates. The scatter plot shows a subset of individuals with low ADOS (blue marker and trend line) and high ADOS (cyan marker and trend line), respectively.

**Figure 4**

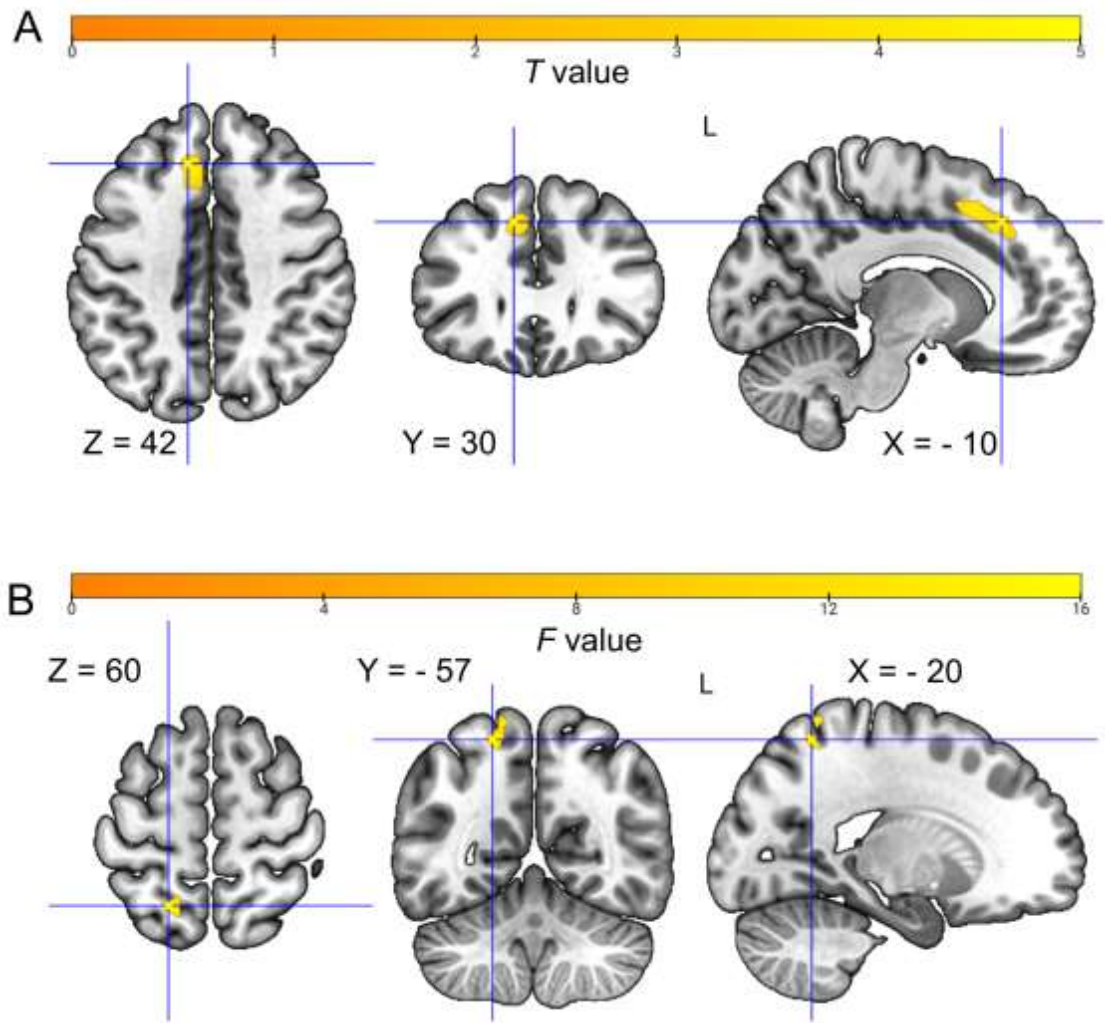

Fig. 4 Whole brain voxel-wise analysis. (A) Difference of rGMV between ASD and TD (ASD < TD) in the left medial superior frontal gyrus. (B) Association between BDI × Diagnoses (ASD and TD) interaction term and rGMV in the left superior parietal gyrus. Thresholded statistical map (*uncorrected p* < 0.001 with extent thresholds 116 and 95 for A and B respectively) overlaid on a high-resolution 1.5 T template image from a single normal subject (MRIcron: ch2.nii.gz) for visualization. Sagittal (X), Coronal (Y), Axial (Z) orientations of the same image and respective peak coordinates.

## Discussion

In this study, we aimed to investigate the association of rGMV of a few a priori ROIs with the inter-individual differences in the severity of co-occurring depression in adults with ASD (18 – 35 years). Furthermore, we investigated whether the rGMV is associated with the interaction between the severity of co-occurring depression and core ASD manifestations. We found the rGMV of the right thalamus to be negatively correlated with the severity of co-occurring depression. Also, the rGMV of left Cerebellum Crus II was associated with the severity of co-occurring depression severity such that the trends were mutually opposite for the subsets of individuals with low and high symptomatic severity of core ASD symptoms, respectively. An exploratory, whole-brain analysis of the differences in rGMV between ASD and age and gender-matched TD individuals remained inconclusive owing to the stringent correction for multiple statistical comparisons.

We found that the GMV of the right thalamus in individuals with ASD was proportionately smaller with increments in the severity of co-occurring depression. While the role of the thalamus in depression is widely reported in the TD population, the findings are slightly contradictory. For example, the GMV of the thalamus is reduced in mild depressive symptoms(53), unipolar depression(54), and major depressive disorder(55–57). An earlier study has even reported a reduction in the rGMV of the right thalamus in otherwise healthy individuals who were cognitively vulnerable to depression(58). However, a meta-analysis has also reported increased thalamic volume to be related to medication-free major depressive disorder(59). Incidentally, the thalamus is also implicated in autism, with notable inconsistencies in the extant literature with regard to the patterns of morphometric changes. Lower Thalamic volumes have been found in individuals with ASD(60–63), but there are also reports of greater thalamic volume(64) and no such difference(65) in the former compared to suitably matched TD. Further, studies have shown the core symptoms of restrictive, repetitive behavior and social cognitive functions in individuals with ASD to be correlated with decreased thalamic volumes(66,67). Alterations in thalamic volume are involved in disruptions of social and emotional processing of information as well as heightened association of the self with negative thoughts and emotions in major depressive disorder (MDD)(68,69). The thalamus acts by collating sensory information and serving as a critical hub for gating this information to the cortex towards modulating the overall levels of activity in the motor, default mode, and central executive networks via reciprocal connections(70–72). This communication affects a range of higher cognitive functions that have been consistently identified to be awry in individuals with ASD that relate to their behavioral symptoms. It is, thus, relevant to study the structural properties of the thalamus in individuals with ASD alongside concurrent (co-occurring) depression to better describe the neural (neuro-morphometric) basis of ASD symptoms. Conspicuously, most of these earlier studies have investigated the neural basis of ASD in isolation (sans their co-occurring) with participants drawn from a wide bracket – 8 to 45 years (68,73–75). Owing to the high prevalence of co-occurring depression (76) and significant changes in brain structure with age in individuals with ASD(77,78), our study here addresses this relatively less explored area. It provides evidence for the relation between thalamic grey matter volume and co-occurring depression in adults with ASD.

We observed a significant interaction between the severity of co-occurring depression (BDI) and clinical manifestations (ADOS) that related to the GMV in the left Cerebellar Crus II in individuals with ASD. While the individuals on clinical severity showed an increasing trend of cerebellar rGMV, the individuals lower on clinical severity had lesser rGMV with proportionate increments of depression severity. Cerebellar abnormalities have been consistently reported to play a significant role in individuals with ASD, with objective brain-based metrics (functional connectivity and gray matter volume) predicting the core symptoms of ASD(79–83). As against the classical motor functions of cerebellum(84), its roles in higher cognitive, social, and emotional functions(85) have come to the fore relatively recently, e.g., contributions to executive function(86,87), language skills / verbal fluency(87), interpreting goal-directed actions of others / social mirroring(88), understanding others' intentions, beliefs, past behaviors, future aspirations, personality traits / social mentalizing(89,90) and emotion attribution from faces(91). The cerebellum, owing to its connections with limbic regions (amygdala, hippocampus, septal nuclei)(92), has a functional role in emotional regulation that is also awry in depression, e.g., increased positive connectivity between Cerebellar Crus II and the temporal poles(93). These anomalies in mentalizing and limited understanding of emotions play an essential role in the maintenance and manifestation of both ASD(81,94) and depression(81,95). With regards to our findings, the Cerebellum Crus II region (located in the posterior lobe of the cerebellum) has been gaining importance for its social cognitive functions (mirroring, mentalizing, facial emotion recognitions) in the ASD population. Most of these studies have focused on children reporting a decrease in the rGMV with increasing clinical severity(81), and a few studies with adults reporting a similar trend(94).

However, earlier studies have dealt with the clinical severity of ASD in isolation. In contrast, we here elucidate the nature of influence severity of clinical symptoms has in interaction with that of co-occurring depression on the volumetric brain phenotype of these individuals. To our knowledge, this, to this day, has not been reported earlier.

Through a brain-wide exploratory analysis, we intended to discover rGMV differences in regions aside from our a priori focus between individuals with ASD and their TD peers that were dependent on depression in both samples. We specifically looked into regions with lesser rGMV in ASD vs. TD minus the influence of depression and the regions whose rGMV varied with the severity of depression differentially between ASD and TD, respectively. These analyses showed brain regions that did not attain the requisite statistical significance (after correction for multiple comparisons). Our finding, especially about the former comparison, agrees with an earlier meta-analysis that could not find any statistically significant differences in rGMV between ASD and TD from 24 studies and > 471 participants in each diagnostic group(96). However, as these brain regions (uncorrected results revealed in our analysis) have been reported in the extant literature about ASD and depression, as described below, they could be considered in future studies. First, we found that the left superior frontal gyrus had a reduced rGMV in ASD compared to TD. The left superior frontal gyrus is a vital structure of the working memory network that is engaged by high-level executive processing and is part of the default-mode network(97). This region was earlier reported to have weak intrinsic functional connections with the posterior cingulate in individuals with ASD that tracked the severity of social cognitive functions(98). Second, we found that the severity of depression was likely associated with the GMV of the left superior parietal gyrus differentially in ASD vs. TD. The superior parietal gyrus, part of the parietal lobe, is also a component of the default-mode network. It generally functions in cognitive tasks, e.g., organization, reward prediction during decision-making, and emotional processing(99). This brain region has recently been shown to bear macromolecular grey matter alterations (as indexed by reduced magnetization transfer ratios) and increased rGMV of a contiguous region - postcentral gyrus in depression(100). Also, a meta-analysis has reported the rGMV of the left superior parietal gyrus to be reduced in the first episode of MDD(101). Importantly, this region has lesser cortical thickness in adults with ASD and, hence, is of relevance in understanding the disorder's manifestations(102).

Finally, we identify a few aspects of our study that may be considered for interpreting our present results as well as for future studies. First, the paucity of phenotypic information regarding depressive co-occurring in the adult age bracket of the ASD and TD participants in the ABIDE database allowed for analyzing only a subset of the participants in our study. In the future, larger datasets replete with the above information may be analyzed to attain greater statistical power and add to our results. This is particularly relevant as a recent study has revealed the existence of neuroanatomical heterogeneity (subtypes) within individuals with ASD(103). Second, the number of females (four) in our ASD sample was too small to analyze the role of gender in influencing the association between rGMV and depression, which could be considered in the future with a larger female and more gender-balanced sample. Third, the BDI is a self-report measure, and owing to various challenges of assessing their own mental states and communication in individuals with ASD (104), it would be prudent to further study the relationship between co-occurring depression and rGMV evident in our results.

Fourth, we chose the stringent statistical threshold of $p < 0.05$ after correcting for multiple comparisons using the FWE method with the additional correction for three a priori ROIs tested in each hemisphere. While this method has a low probability of false positives, it does have a greater probability of false negatives (wrongly indicating the absence of an effect) in that it may fail to identify the rGMV changes in the brain(105). This may be addressed by adopting alternative statistical methods, e.g., False Discovery Rate (FDR), while reporting the analyses(106).

In conclusion, our results demonstrate that within adults with ASD, the rGMV of the right thalamus is negatively correlated with the severity of co-occurring depression. Furthermore, the pattern of correlation between the severity of co-occurring depression and rGMV of left cerebellum crus II is mutually opposite between the two subsets of individuals with low and high clinical severity of ASD, respectively. These results add to the existing body of literature by addressing an interplay of prevalent co-occurring depression and the core deficits of ASD regarding the structural heterogeneity of the brain, specifically in adults with ASD. It underscores the need for further exploration of the brain in adults with ASD and co-occurring psychopathologies with larger samples to uncover improved neuroanatomical biomarkers. The findings are of relevance in expounding the neurobiological underpinnings and in identifying structural neuroimaging-based biomarkers of ASD.

**Author contributions**

DD: investigation, data curation. MaC: investigation, writing, proof-reading, data curation. SS: investigation. CM: analysis, writing. U: writing. CS: investigation, data curation. MrC: investigation, analysis, writing, funding acquisition, project administration. All authors contributed to the article and approved the submitted version.

**Conflicts of Interest**

The authors declare no conflicts of interest.

**Acknowledgments**

The research was funded by the Research Initiation Grant by IIIT-D as well as by a grant from the Center for Design and New Media (A TCS Foundation Initiative supported by the Tata Consultancy Services) at IIIT-Delhi to Dr. Mrinmoy Chakrabarty. The funders were not involved in the study design, collection, analysis, interpretation of data, the writing of this article, or the decision to submit it for publication.

# References

1. Klin A, Jones W, Schultz R, Volkmar F, Cohen D. Defining and Quantifying the Social Phenotype in Autism. American Journal of Psychiatry. 2002 Jun;159(6):895–908.
2. Mannion A, Brahm M, Leader G. Comorbid Psychopathology in Autism Spectrum Disorder. Rev J Autism Dev Disord. 2014;1(2):124–34.
3. Lainhart JE. Psychiatric problems in individuals with autism, their parents and siblings. Vol. 11, International Review of Psychiatry. 1999. p. 278–98.
4. Al-Beltagi M. Autism medical comorbidities. World J Clin Pediatr. 2021;10(3):15–28.
5. Simonoff E, Pickles A, Charman T, Chandler S, Loucas T, Baird G. Psychiatric disorders in children with autism spectrum disorders: Prevalence, comorbidity, and associated factors in a population-derived sample. J Am Acad Child Adolesc Psychiatry. 2008;47(8):921–9.
6. Simonoff E, Jones CRG, Baird G, Pickles A, Happé F, Charman T. The persistence and stability of psychiatric problems in adolescents with autism spectrum disorders. J Child Psychol Psychiatry. 2013;54(2):186–94.
7. Sobanski E, Brüggemann D, Alm B, Kern S, Deschner M, Schubert T, et al. Psychiatric comorbidity and functional impairment in a clinically referred sample of adults with attention-deficit/hyperactivity disorder (ADHD). Eur Arch Psychiatry Clin Neurosci. 2007;257(7):371–7.
8. Lugnegård T, Hallerbäck MU, Gillberg C. Personality disorders and autism spectrum disorders: What are the connections? Compr Psychiatry. 2012;53(4):333–40.
9. Joshi G, Wozniak J, Petty C, Martelon MK, Fried R, Bolfek A, et al. Psychiatric comorbidity and functioning in a clinically referred population of adults with autism spectrum disorders: A comparative study. J Autism Dev Disord. 2013;43(6):1314–25.
10. Matson JL, Cervantes PE. Commonly studied comorbid psychopathologies among persons with autism spectrum disorder. Vol. 35, Research in Developmental Disabilities. 2014. p. 952–62.
11. Mayes SD, Calhoun SL, Murray MJ, Zahid J. Variables Associated with Anxiety and Depression in Children with Autism. J Dev Phys Disabil. 2011;23(4):325–37.
12. Ghaziuddin M, Ghaziuddin N, Greden J. Depression in Persons with Autism: Implications for Research and Clinical Care. Vol. 32, Journal of Autism and Developmental Disorders. 2002. p. 299–306.
13. Antshel KM, Russo N. Autism Spectrum Disorders and ADHD: Overlapping Phenomenology, Diagnostic Issues, and Treatment Considerations. Vol. 21, Current Psychiatry Reports. 2019.
14. Postorino V, Kerns CM, Vivanti G, Bradshaw J, Siracusano M, Mazzone L. Anxiety Disorders and Obsessive-Compulsive Disorder in Individuals with Autism Spectrum Disorder. Vol. 19, Current Psychiatry Reports. 2017.
15. Kim JA, Szatmari P, Bryson SE, Streiner DL, Wilson FJ. The prevalence of anxiety and mood problems among children with autism and Asperger syndrome. Autism. 2000;4(2):117–32.
16. Chandrasekhar T, Sikich L. Challenges in the diagnosis and treatment of depression in autism spectrum disorders across the lifespan. Dialogues Clin Neurosci. 2015;17(2):219–27.
17. Hollocks MJ, Lerh JW, Magiati I, Meiser-Stedman R, Brugha TS. Anxiety and depression in adults with autism spectrum disorder: A systematic review and meta-analysis. Vol. 49, Psychological Medicine. 2019. p. 559–72.

# Supplement

Gray matter volume correlates with Co-occurring Depression in Autism Spectrum Disorder.

Authors: Dolcy Dhar[1#], Manasi Chaturvedi[1,2#], Saanvi Sehwag[1], Chehak Malhotra[3], Udit[4], Chetan Saraf[5], Mrinmoy Chakrabarty[1,2#*]

[1]Dept. of Social Sciences and Humanities, Indraprastha Institute of Information Technology Delhi, New Delhi 110020, INDIA.

[2]Centre for Design and New Media, Indraprastha Institute of Information Technology Delhi, New Delhi 110020, INDIA.

[3]Dept. of Mathematics, Indraprastha Institute of Information Technology Delhi, New Delhi 110020, INDIA.

[4]Dept. of Computational Biology, Indraprastha Institute of Information Technology Delhi, New Delhi 110020, INDIA.

[5]Dept. of Computer Science and Engineering, Indraprastha Institute of Information Technology Delhi, New Delhi 110020, INDIA.

**# equal contribution (Co-first author)**

Correspondence (*):

Mrinmoy Chakrabarty, Ph.D.

Assistant Professor (Cognitive Science), Dept. of Social Sciences and Humanities, Indraprastha Institute of Information Technology Delhi (IIITD), New Delhi 110020, INDIA. TEL: +91-011-26907-363; E-mail: mrinmoy@iiitd.ac.in

*Data Pre-processing*

All imaging data were processed and analyzed using a combination of Statistical Parametric Mapping 12 (v7771)(1) + Computational Anatomy Toolbox v12.8.1 (r2043) (2) hosted on MATLAB 2022a. All scans were approximately aligned such that their anterior commissures coincided with those of the default template. This was followed by implementing the default automated segmentation pipeline of the CAT12 toolbox. We took the default approach employed by CAT12: the Adaptive Maximum A Posterior (AMAP) approach, where prior knowledge of tissue probabilities is not always required. An initial segmentation into three pure classes (grey matter, white matter, and cerebrospinal fluid) was performed. This was followed by a Partial Volume Estimation (PVE) of two additional classes: GM-WM and GM-CSF. For this, all the T1-weighted scans were first registered to a standard template for affine regularization and were skull-stripped. This was followed by spatially registering all the scans to a standardized template in Montreal Neurological Institute (MNI) space, normalizing, and segmenting in different tissue volumes (grey matter, white matter, and CSF). Segmented GM normalized bias-corrected volumes from each scanning center were run through an independent "Sample Homogeneity" data quality check. Individual IQRs for the three datasets obtained were as follows: Barrow's Neurological Institute: 0.873 – 0.896; Indiana University: 0.872 – 0.888; Olin's Neuropsychiatric Research Center: 0.859 – 0.879. The total intracranial volumes (TIV) for all subjects were then calculated (ASD [mean ± SD] = 1.58 ± 0.14 cm$^3$;

TD = 1.56 ± 0.14 cm$^3$), and finally, the grey matter segmented tissue probability maps were smoothed using an 8 mm full-width half-max (FWHM) Gaussian kernel. Here, morphometric analyses objectively assess regional cortical and subcortical volumes across various brain morphologies. It employs a series of algorithms to determine volumetric differences in structural MR images spatially and enables researchers to utilize statistical methods for precisely studying the brain-behavior relationship.

*Defining a priori Regions of Interest*

We narrowed down on a priori regions of interest for the analysis as the necessity for a severe correction for type I error at the whole brain level, especially with a relatively smaller sample size, may limit the statistical power(3). Accordingly, a literature search was conducted for whole-brain, volumetric gray matter MRI studies (excluding white matter analysis and diffusion tensor imaging) involving adults with ASD (18-35 years; the age bracket of our ABIDE-II samples) using the Sleuth application of the BrainMap repository(4) and PUBMED. Keywords selected for the search were a few salient cognitive issues in individuals with ASD that overlapped with depression as well. These issues have been linked with the clinical manifestations of these disorders separately. The reason was to narrow down brain regions that related to the same set of issues of both ASD and depression so as to investigate the relationship of their rGMV in ASD with co-occurring depression. The timeline of the search was kept between 2004 – 2024. First, a search was conducted with keywords – affect sensory-motor integration, social cognition, theory of mind, motor impairment, and (co-occurring) depression. This returned no results. Next, a search was conducted to look for studies only in ASD adults with all the above particulars except (co-occurring) depression, which also resulted in no studies. Finally, another search was conducted to look for studies in typically-developed adults with depression with the above particulars, which yielded no results. Thus, we depended on earlier systematic reviews and meta-analyses to choose three regions of a priori interest that bore relevance to the above cognitive functions and had prior evidence of gray matter volume anomalies in both ASD and depression to investigate the relationship of the volumes of those structures in ASD adults with co-occurring depression. Thus, we chose a) the Anterior Cingulate Cortex as it is plays a crucial role in social cognition and consistently implicated in ASD (5,6) and depression (7,8) and b) the Thalamus for its role in sensory motor integration, emotional regulation, consciousness which are impaired in ASD and depression with structural anomalies reported in these two clinical cohorts (9–17) and c) Cerebellum for its role in motor coordination, emotions and theory of mind (18–22) and specifically, the Crus II of Cerebellum which has been implicated in both ASD and depression (23). Finally, anatomical brain ROIs were defined by the WFU PickAtlas Toolbox for SPM version 12 using masks created from the Automated Anatomical Labelling(24–26). A predefined anatomic label was applied to identify ROIs relevant to this study for the left and right hemispheres separately. See Fig. 1 for the spatial locations of the defined ROIs.